\documentclass[aps,preprint,showpacs,amsmath,amssymb,epsfig]{revtex4}
\usepackage{graphicx}
\usepackage[usenames]{color}
\usepackage{epstopdf}
\usepackage{bm}

\usepackage[colorlinks=true, urlcolor=navyblue, linkcolor=navyblue, citecolor=navyblue]{hyperref}
\definecolor{navyblue}{rgb}{0,0.08,0.45}
\begin{document}

\preprint{SLAC-PUB-16078}

\title{Connecting the Hadron Mass Scale to the Fundamental Mass Scale of Quantum Chromodynamics} 

\author{
A. Deur,$^{\njlab}$
S. J. Brodsky,$^{\nslac}$
G. F. de Teramond.$^{\nucr}$
}

\affiliation{
\baselineskip 2 pt
\centerline{{$^{\njlab}$Thomas Jefferson National Accelerator Facility, Newport
News, VA 23606, USA}}
\centerline{{$^{\nslac}$SLAC National Accelerator Laboratory, Stanford University,
Stanford, California 94309, USA}}
\centerline{{$^{\nucr}$Universidad de Costa Rica, San Jos\'e, Costa Rica}}
}

\newcommand{\njlab}{1}
\newcommand{\nslac}{2}
\newcommand{\nucr}{3}

\begin{abstract}
Establishing an explicit connection between the long distance physics of confinement
and the dynamical interactions of quarks and gluons at short distances
has been a long-sought goal of quantum chromodynamics. 
Using holographic QCD, we derive a direct analytic  relation between the scale $\kappa$ which determines the masses of hadrons and the scale $\Lambda_{s}$ which controls 
the predictions of perturbative QCD at very short distances. The resulting
prediction $\Lambda_{s}=0.341\pm0.032$ GeV in the $ \overline{MS}$
scheme agrees well with the experimental average $0.339\pm0.016$
GeV. We also derive a relation between $\Lambda_{s}$ and the QCD string tension $\sigma$. 
This connection between the fundamental hadronic scale underlying the physics of
quark confinement and the perturbative QCD scale controlling hard collisions can be carried out in any renormalization scheme.

\end{abstract}


\maketitle

\section{Introduction}

Quantum Chromodynamics (QCD) provides a fundamental description of the dynamics 
binding quarks and gluons into hadrons. QCD is well understood
at high momentum transfer where perturbative calculations are applicable.
Establishing an explicit relation between the short-distance regime and 
the  large-distance physics of color confinement has been a long-sought goal.
A major challenge is to relate the parameter $\Lambda_{s}$, which controls the predictions of 
perturbative QCD (pQCD) at short distances, to the masses of hadrons or to the QCD string 
tension $\sigma$.  In this paper, we shall show how theoretical insights into 
color confinement and hadron dynamics derived from holographic QCD at large distances
lead to an analytical relation between hadronic masses and $\Lambda_{s}$.  The resulting
prediction,  $\Lambda_{s}=0.341\pm 0.032$ GeV, as defined in the $\overline{MS}$ scheme, agrees well with
the experimental value $0.339\pm 0.016$ GeV~\cite{PDG:2014}.   In addition, our value for $\sigma$,  
$0.191\pm 0.009$ GeV$^2$ is in excellent agreement with the phenomenological value $\sigma \simeq 1$ GeV/fm $=0.197$ GeV$^2$~\cite{Bali:2013kia}.  Conversely, the experimental value of $\Lambda_s$  obtained from measurements at high momentum transfer can be used to predict the masses of hadrons.

The masses of hadrons such as the proton and $\rho$ meson must emerge 
from the fundamental forces of QCD which confine their quark constituents.   Naively, one would expect the hadronic mass scale of the order of a GeV  to be explicitly 
present in the QCD Lagrangian.  
However,
the only scale appearing in the QCD Lagrangian  for hadrons  made of light quarks corresponds to quark masses of the order of a few MeV, too small to be relevant.  
An important mass scale, $\Lambda_s$,  does exist, however, when one quantizes the theory. This parameter controls the strength of the coupling of quarks in the asymptotic freedom domain where  quarks interact at short distances.  The  explicit definition of $\Lambda_s$ depends on the renormalization scheme used to regulate the ultraviolet divergences of the perturbative theory.   The connection between $\Lambda_s$ and the mass scale which controls confinement  in a scale-invariant field theory is called ``dimensional transmutation";   this mechanism is assumed to originate from the renormalization group equations of the underlying quantum theory~\cite{Gross:1973id, Politzer:1973fx, Dvali:2011uu}.

This paper will present a new  systematic approach which  analytically links $\Lambda_s$ to hadron masses.  It will allow us to precisely predict the
 value of  $\Lambda_s$ taking a hadronic mass as input, or, conversely,  to calculate the  hadron masses using 
 $\Lambda_s$.    Another mass scale, relevant to confinement, is the string tension $\sigma$, which 
determines the hadron mass spectrum and the Regge slopes based on a model utilizing a static quark-quark potential.

We will utilize the value of $\Lambda_s$ as 
defined using the  ${\overline{MS}}$  renormalization scheme, although our results can be implemented for any choice of the renormalization 
procedure. The  parameter $\Lambda_s $ 
can be determined to high precision from experimental measurements of  high-energy, short-distance,  processes where the strength of 
QCD  is small because of asymptotic freedom~\cite{Gross:1973id, Politzer:1973fx},  and pQCD is thus applicable.  The value of $\Lambda_s$ can also be 
determined to high accuracy using numerical  lattice techniques~\cite{Aoki:2013ldr}; it can also be predicted from the pion decay constant $F_{\pi}$ using Optimized Perturbation Theory~\cite{Kneur}.

We will use a semiclassical approximation 
to  QCD in its large-distance regime which follows from the connections between light-front dynamics and  its 
holographic mapping to higher-dimensional anti-de Sitter (AdS$_5$) space-time. AdS$_5$ is a mathematical  construction which provides an elegant geometric representation of the conformal group.

In holographic QCD -- often referred to as  ``AdS/QCD" -- the forces that bind and confine quarks are derived from the 
``soft-wall" modification of the geometry in the  fifth dimension $z$ of AdS$_5$  space~\cite{Karch:2006pv}. 
The specific modification of the AdS$_5$ action, a dilaton factor $e^{\kappa^2 z^2}$,
leads to Regge trajectories and is compatible with light-front confinement dynamics~\cite{deTeramond:2013it}. This form of the dilaton factor can be connected to a basic mechanism due to de Alfaro, Fubini and 
Furlan~\cite{Brodsky:2013ar, deAlfaro:1976je}, which allows for the emergence of a mass scale  $\kappa$ in the equations of motion and the Hamiltonian of  the theory
while conserving the conformal invariance of the action.  The soft-wall modification of AdS$_5$  space leads directly to the form of the 
quark-confining light-front potential, namely a harmonic oscillator potential.  
The scale $\kappa$  controlling quark confinement  also predicts the hadron masses.
For example, $\kappa$ can be determined from the   $\rho$ hadron mass: $\kappa = M_\rho/\sqrt{2}= 0.548 $ GeV~\cite{Brodsky:2014yha}.
In the case of heavy quarks, the light-front harmonic oscillator potential transforms to a linear potential in a nonrelativistic Schr\"odinger equation characterized
by the string tension $\sigma=2\kappa^2/\pi$~\cite{Trawinski:2014msa}. 
This  approach to hadronic physics and color confinement, called ``Light-Front Holographic QCD"~\cite{Brodsky:2014yha} 
and its superconformal extension~\cite{deTeramond:2014asa,  Dosch:2015nwa} can explain many hadronic properties 
of the light mesons
and baryons,  such as the observed mass pattern of  radial and orbital excitations.  In addition,  the application of superconformal algebra leads to supersymmetric relations between mesons and baryons with internal orbital angular momentum satisfying $L_M= L_B+1$, which can be extended to heavy hadrons~\cite{Dosch:2015bca}.  
Holographic QCD also predicts the light-front wavefunctions which control form factors, transverse momentum distributions, and other dynamical features of hadrons.

The essential feature of Light-Front Holographic QCD which we shall utilize in this paper  is the fact that it prescribes the form of the QCD coupling $\alpha_s(Q^2)$ in the  nonperturbative domain~\cite{Brodsky:2010ur}.  
($Q^2$ is the scale at which the hadron is probed. It is defined as the absolute value of the square of the 4-momentum transferred by the scattered electron to the nucleon)
On the other hand, the small-distance 
physics where asymptotic freedom rules,  is well-described by pQCD. The two regimes overlap at intermediate distances, a phenomenon called 
``quark-hadron duality"~\cite{Bloom:1970xb}. This duality will permit us to match the hadronic and partonic
descriptions and obtain an analytical relation between $\Lambda_s$  and hadron masses.  

We shall relate the long and short-distance scales by matching the AdS/QCD form of the QCD running coupling $\alpha_{s}(Q^2)$ at low $Q^2$, which depends on 
$\kappa$, to the pQCD form of the coupling, which explicitly depends on $\Lambda_{\overline {MS}}$.  
In pQCD, the $Q^2$-dependence of $\alpha_{s}(Q^2)$ originates from short-distance quantum 
effects  which  are folded into its definition;  the scale $\Lambda_{\overline {MS}}$  controls this space-time 
dependence~\cite{Gross:1973id,Politzer:1973fx}.   
Analogously, the space-time dependence of the AdS/QCD 
coupling derives from the dilaton modification of the AdS space-time curvature which depends on $\kappa$~\cite{Brodsky:2010ur}.

\section{The Effective Charge $\alpha_{g_1}(Q^2)$}

As Grunberg~\cite{Grunberg:1980ja}  has emphasized, it is natural to define the QCD coupling from a physical observable which is perturbatively calculable at large $Q^2$. 
This is analogous to QED,  where the standard running Gell Mann-Low coupling $\alpha$ is defined from the elastic scattering amplitude for heavy leptons.  A physically defined  ``effective charge" incorporates nonperturbative dynamics at low scales, and it evolves at high scales to the familiar pQCD form $4\pi / \beta_0 \log\left(Q^2/\Lambda_s^2\right)$, as required by asymptotic freedom at high scales.   As expected on physical grounds, effective charges are finite and smooth at small $Q^2$.

We will focus  on  $\alpha_{g_1}(Q^2)$ which is the best-measured effective charge~\cite{Deur:2005cf}. The effective coupling is defined from the  Bjorken sum rule~\cite{Bjorken:1966jh}:  
\begin{equation} \label{BjSR}
\frac{\alpha_{g_{1}}(Q^2)}{\pi} = 1 -  \frac{6}{g_{A}} \int_{0}^{1} dx \, g_{1}^{p-n}(x,Q^{2}),
\end{equation}
where $x$ is the Bjorken scaling variable, $g_1^{p-n}$  is the isovector component of the nucleon first spin structure
function and $g_A$ is the nucleon axial charge. The effective charge $\alpha_{g_1}(Q^2)$ is kinematically constrained
 to satisfy $\alpha_{g_1}\left(Q^2=0\right)=\pi$. The Gerasimov-Drell-Hearn sum rule~\cite{Drell:1966jv}  
 implies that $\alpha_{g_1}(Q^2)$ is nearly conformal in the low-$Q^2$ domain~\cite{Deur:2005cf}. 
The coupling $\alpha_{g_1}(Q^2)$ plays a role analogous 
to the Gell-Mann-Low coupling $\alpha(Q)$ of QED~\cite{Brodsky:2010ur}. 
The $V$ scheme defined from the heavy quark potential is not normally used as an effective charge because of the presence 
of infrared divergences in its pQCD expansion, divergences which can be controlled  by color confinement~\cite{the:gluon H graph divergence}.

\begin{figure}
\centerline{\includegraphics[width=.45\textwidth]{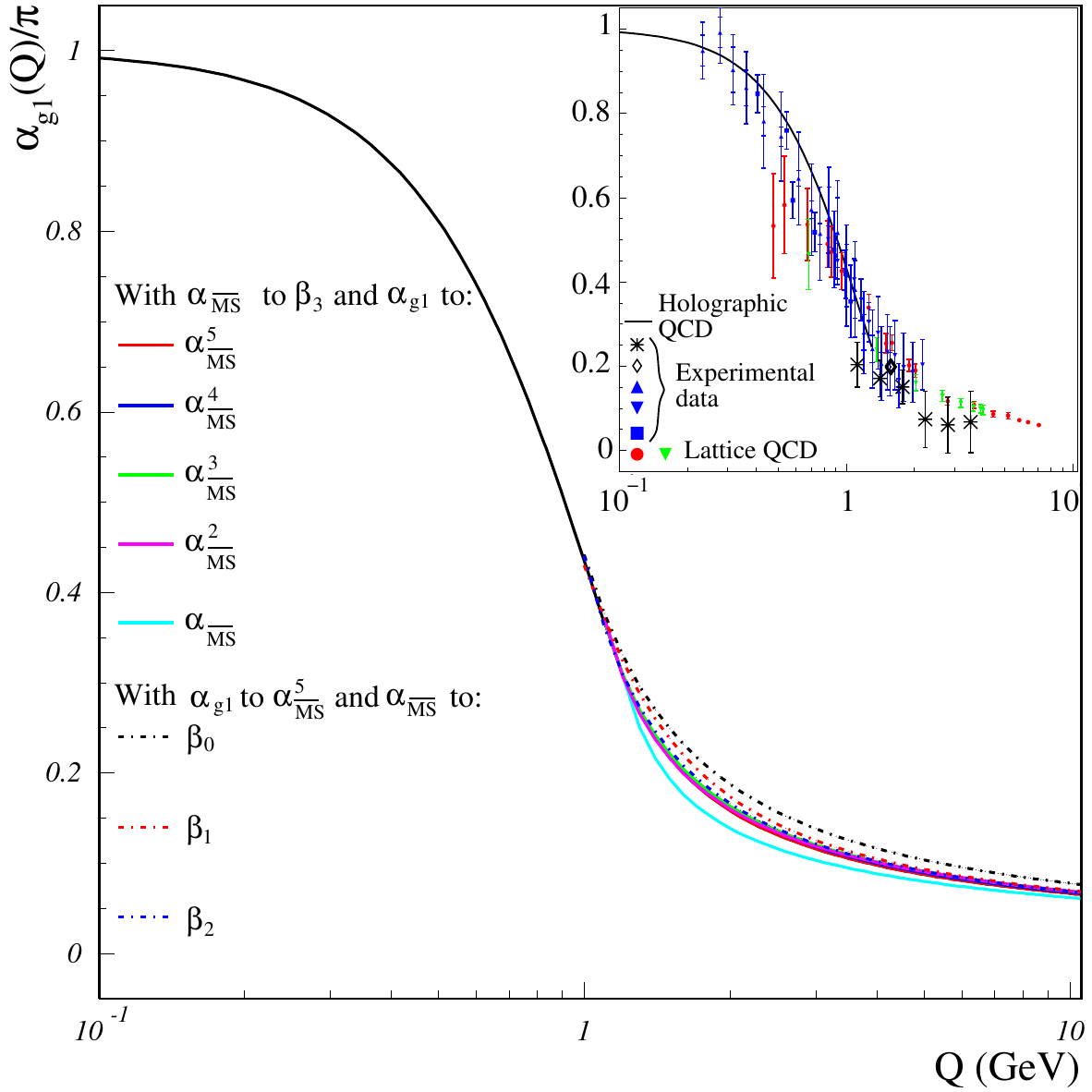}}
\caption{\label{Fig:order dependence}The dependence of $\alpha_{g_1}$ on the orders of the $\beta$ and 
$\alpha_{\overline{MS}}$ 
series.  The continuous black line is the AdS coupling. The continuous colored lines are the 
matched pQCD couplings for all available orders in the $\alpha_{\overline{MS}}$ series (the order of the $\beta$ series 
was kept at $\beta_{3}$). The dash-dotted colored lines are the matched couplings at different orders in the 
$\beta$ series (the order of the series was kept at $\alpha_{\overline{MS}}^{5}$).   The curves beyond the leading order
are observed to be remarkably close. The comparison between the AdS coupling and the data is shown in the embedded figure. 
This comparison is  shown within the  range of validity of holographic QCD.} 
\end{figure}

Light-front holographic QCD predicts the behavior of  $\alpha_{g_1}(Q^2)$ at  small values of $Q^2$. The physical coupling measured at the scale $Q^2$ is the two-dimensional Fourier transform of the light-front transverse coupling~\cite{Brodsky:2010ur}:
\begin{equation} \label{alphaAdS}
\alpha^{AdS}_{g_1}(Q^2) = \pi \exp{\left(-Q^2/4 \kappa^2\right)}.
\end{equation}
Eq.~(\ref{alphaAdS}) explicitly connects the small-$Q^2$ dependence of $\alpha_{g_1}(Q^2)$ to $\kappa$, and thus  to hadronic masses.
It is valid only at small $Q^2$ where QCD is a strongly coupled theory with a nearly conformal behavior, and thus where the  holographic QCD methods are applicable.
The behavior of the running coupling predicted by AdS/QCD is in remarkable agreement with the experimental 
measurements~\cite{Deur:2005cf} as seen in the inset of 
Fig.~\ref{Fig:order dependence}. Even though there are no free parameters since $\kappa$ is fixed by the hadron 
masses, the predicted Gaussian shape of  $\alpha^{AdS}_{g_1}(Q^2)$ agrees very well with the data.

The large $Q$-dependence of $\alpha_{s}$ is computed from the renormalization group equation 
\begin{equation} \label{beta}
Q^2{d \alpha_{s}}/{dQ^2} =\beta(Q^2)=-(\beta_0 \alpha_s^2 + \beta_1 \alpha_s^3 + \beta_2 \alpha_s^4 +  \cdots),
\end{equation}
where the $\beta_i$ coefficients are known up to $\beta_3$  in the $\overline{MS}$ scheme~\cite{PDG:2014}.
Furthermore, 
$\alpha_{g_1}^{pQCD}(Q^2)$ can be itself expressed as a perturbative expansion in 
$\alpha_{\overline{MS}}(Q^2)$.
Thus pQCD predicts the form of  $\alpha_{g_1}(Q^2)$ at large $Q^2$:
\begin{equation}  
\label{eq: alpha_g1 from Bj SR}
\alpha_{g_{1}}^{pQCD}(Q^2)  = \pi \Big[{\alpha_{\overline{MS}}}/{\pi} +  
a_1\left({\alpha_{\overline{MS}}}/{\pi}\right)^{2} + \\ a_2\left({\alpha_{\overline{MS}}}/{\pi}\right)^{3} +  \cdots \Big].
\end{equation} 
The coefficients $a_i$ are known up to order $a_3$~\cite{Baikov:2010je}.

\begin{figure}
\centerline{\includegraphics[width=.45\textwidth]{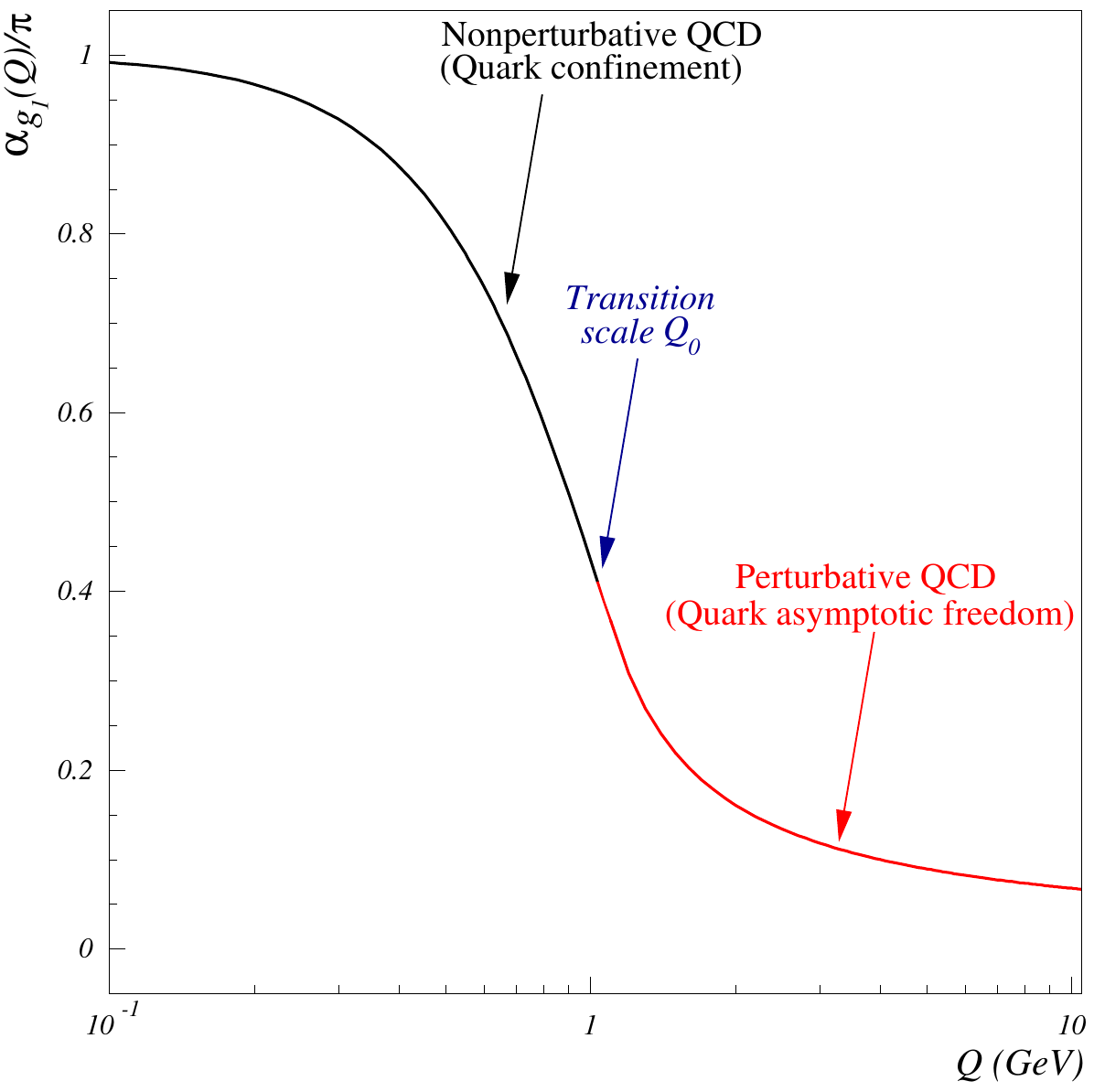}}
\caption{\label{Fig:matching} Unified strong coupling from the analytic matching of perturbative and nonperturbative QCD regimes. 
The analytic matching determines the relation between $\Lambda_{\overline {MS}}$ and hadron masses as well as the transition scale $Q_0$
interpolating between the large and short-distance regimes of QCD.}
\end{figure}

The dependence of $\alpha_{g_1}$ on  $Q^2$  must be analytic.  The existence at moderate values of $Q^2$ of a dual description 
of QCD in terms of either quarks and gluons or hadrons (``parton-hadron duality"~\cite{Bloom:1970xb}) implies that the AdS/QCD
and pQCD forms, Eqs.~\ref{alphaAdS} and~\ref{eq: alpha_g1 from Bj SR} can be matched. This can be done by  imposing continuity 
of both $\alpha_{g_1}(Q^2)$ and its derivative, as shown in Fig.~\ref{Fig:matching}. The  resulting two equalities then provide a unique value of 
$\Lambda_s$ from  the scheme-independent scale $\kappa$, and fix the scale $Q_0$ 
characterizing the transition between the large and short-distance regimes of QCD.

We have solved the two-equation system resulting from the matching of the two $\alpha_{g_1}(Q^2)$ and their derivatives.  This is done analytically at leading order of Eqs.~\ref{beta} 
and~\ref{eq: alpha_g1 from Bj SR}, and numerically up to fourth order. The leading-order analytical relation between $M_\rho= \sqrt{2} \kappa $ and $\Lambda_{\overline{MS}}$ is:
\begin{equation}  \label{eq: Lambda LO analytical relation}
\Lambda_{\overline{MS}}=M_\rho e^{-a}/\sqrt{a},
\end{equation} 
with $a=4\big(\sqrt{ln(2)^{2}+1+\beta_{0}/4}-ln(2)\big)/\beta_{0}$. For $n_{f} = 3$ quark flavors, $a\simeq 0.55$.

\begin{figure}
\centerline{\includegraphics[width=.45\textwidth]{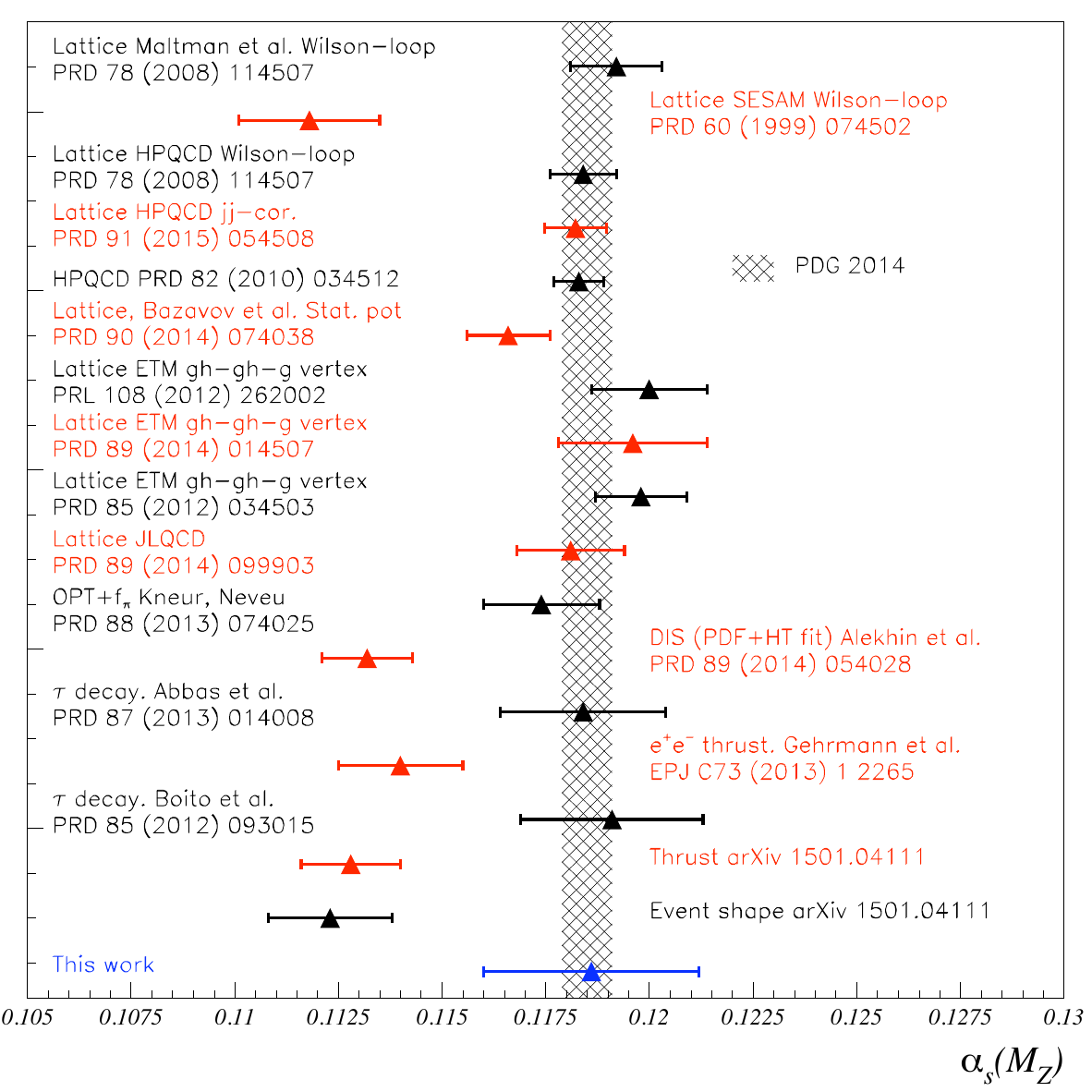}}
\caption{\label{Fig:comparison} Comparison between our result and  determinations of $\alpha_{\overline{MS}}(M_Z)$ from the high precision experimental and lattice measurements. The world average~\cite{PDG:2014} is shown as the vertical band.}
\end{figure}

Since the value of $Q_0$ is relatively small, higher orders in perturbation theory are essential for obtaining an accurate relation 
between $\Lambda_s$  and hadron masses, and to evaluate the convergence of the result. 
In  Fig.~\ref{Fig:order dependence} we show how  $\alpha^{pQCD}_{g_{1}}(Q^2)$ depends on the $\beta_{n}$ and 
$\alpha_{\overline{MS}}$ orders used in Eqs.~(\ref{beta}) and~(\ref{eq: alpha_g1 from Bj SR}), respectively.  The curves converge  quickly to a 
universal shape independent of the perturbative order;   at order $\beta_{n}$ or $\alpha_{\overline{MS}}^{n}$, $n > 1$, the
$\alpha^{pQCD}_{g_{1}}(Q^2)$ are nearly identical. 
Our result at $\beta_3$, the same order to which the experimental value of $\Lambda_{\overline{MS}}$ is 
extracted, is $\Lambda_{\overline{MS}}=0.341 \pm 0.032$ GeV for $n_{f} = 3$. The uncertainty stems
from the extraction of $\kappa$  from the $\rho$ or  proton mass ($\pm 0.024$), the truncation uncertainty in Eq.~(\ref{eq: alpha_g1 from Bj SR}) ($\pm 0.021$) and the 
uncertainty from  the chiral limit  extraction of $\kappa$ ($\pm \, 0.003$ GeV). Our uncertainty is competitive with that of the individual experimental determinations,  
which combine to $\Lambda_{\overline{MS}} = 0.339 \pm 0.016$ GeV~\cite{PDG:2014}.  Including results from numerical lattice techniques, which provide 
the most accurate determinations of $\Lambda_{\overline{MS}}$, the combined world average is $0.340 \pm 0.008$ GeV~\cite{PDG:2014}. We show in 
Fig.~\ref{Fig:comparison} how our calculation compares with this average, as well as with recent lattice results and the best experimental determinations.

Our relation can also be expressed in term of the string tension $\sigma$. 
At LO we have the analytical relation:
\begin{equation}
\sigma = a e^{2a} \Lambda^2_{\overline{MS}}/ \pi .
\end{equation}
The numerical relation at orders $\beta^3$ and $\alpha_{\overline{MS}}^{4}$ of Eqs.~(\ref{beta}) and~(\ref{eq: alpha_g1 from Bj SR}), respectively, yields $\sigma = 1.655 \Lambda^2_{\overline{MS}}= 0.191 \pm 0.009$ 
GeV$^2$ for $\Lambda_{\overline{MS}}=0.340 \pm 0.008$ GeV, in excellent agreement  with the determination from phenomenology.

Our holographic QCD approach also determines the transition scale $Q_0$.  We can interpret $Q_0$ as the effective
initial scale where  DGLAP~\cite{DGLAP} and ERBL~\cite{ERBL} evolutions begin. The scale $Q_0$ also sets the limit of validity of holographic QCD and how  it breaks down as one
 approaches  the pQCD domain. 
 At order $\beta_0$, we have:
\begin{equation}  \label{eq: Q_0 LO analytical relation}
Q_{0}=M_\rho/\sqrt{a}.
\end{equation} 
At order $\beta_3$,  $Q_0^2 \simeq 1.25 \pm 0.19$ GeV$^2$. This value is similar to the traditional lower 
limit $Q^2 > 1$ GeV$^2$ used for pQCD. 
An approximate value similar to ours was found in Ref.~\cite{Courtoy:2013qca}, which terminates the evolution of 
$\alpha_s(Q^2)$ near $Q^2 \simeq 1$ GeV$^2$ in order to enforce  parton-hadron 
duality for the proton structure function $F_2(x,Q^2)$  measured in deep-inelastic experiments.

\begin{figure}
\centerline{\includegraphics[width=.45\textwidth]{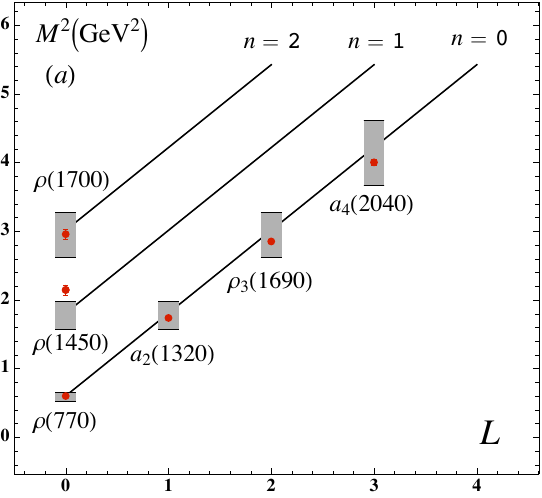} \hspace{10pt}
\includegraphics[width=.45\textwidth]{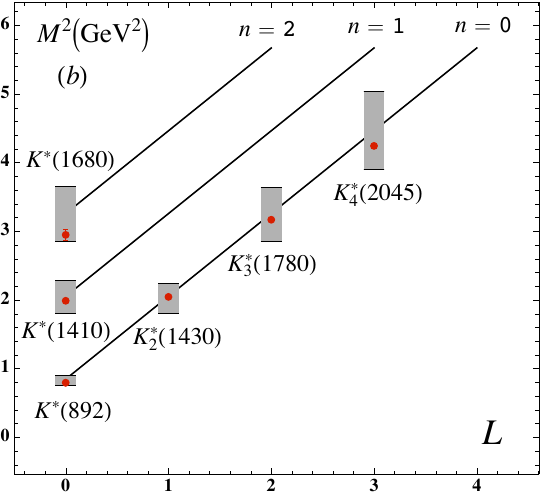}}
\caption{\label{Fig:masses} The predicted mass spectrum for  the light vector mesons as a function of the internal orbital angular momentum $L$  and the radial excitation $n$:  (a)  unflavored mesons and (b)  strange mesons. The red dots are the experimental values. The dark lines represent  our mass determination and the gray bands the uncertainty. The only parameter entering this determination is the world average $\Lambda_{\overline{MS}} = 0.340 \pm 0.008$ GeV and, in addition for the strange mesons, the strange quark mass~\cite{Brodsky:2014yha}. The decay widths of the mesons are not accounted for in the calculation.}
\end{figure}

Conversely,  we can use the ratio between $\Lambda_{\overline {MS}}$ and $\kappa$ to predict the hadron spectrum. For example, starting with the measured
value of $\Lambda_{\overline {MS}}$, $0.340 \pm 0.008$ GeV~\cite{PDG:2014}, one obtains $M_{\rho}=0.777 \pm  0.051$ GeV, in near perfect agreement 
with the measurement $M_{\rho}=0.775 \pm  0.000 $ GeV \cite{PDG:2014}.  The values for the uncertainty comes from the following sources: 0.045 GeV from the truncation of the series, Eq.~(\ref{eq: alpha_g1 from Bj SR}), 0.021 GeV from the uncertainty on $\Lambda_{\overline {MS}}$~\cite{PDG:2014} and 0.009 GeV from the truncation of the $\beta$ series, Eq.~(\ref{beta}). Our computed proton or neutron mass, $M_N=1.092 \pm 0.073$ 
GeV, is 2$\sigma$ higher than the averaged experimental values, $0.939 \pm 0.000$ GeV.
Other hadron masses are calculated as orbital and radial excitations of the hadronic Regge trajectories~\cite{Brodsky:2014yha, deTeramond:2013it} 
Thus, using $\Lambda_{\overline {MS}}$ as the only input, the hadron mass spectrum is calculated 
self-consistently within the holographic QCD framework, as shown in Fig.~\ref{Fig:masses} for the vector mesons. 
We emphasize that QCD has no knowledge of conventional units of mass such as GeV; only ratios are predicted. 
Consequently our work essentially predicts the ratios $\Lambda_{\overline {MS}}/M$ where  $M$ is any hadron mass. For the same reason,  the ratio $\Lambda_{\overline {MS}}/F_{\pi}$ is computed in Ref.~\cite{Kneur}.

\section{Conclusions}

In summary, we have obtained an explicit relation
between the quark-confining nonperturbative dynamics of QCD at  large-distances based on the semiclassical light-front holographic approximation of QCD 
and the 
short-distance dynamics of perturbative QCD. The analytic form of the 
QCD running coupling at all energy scales is also determined.   
The result is an explicit link
of the perturbative QCD scale $\Lambda_{\overline {MS}}$ to the masses of the observed hadrons.
The predicted value 
$\Lambda_{\overline{MS}} = 0.341 \pm 0.032$ GeV agrees well with the experimental average 
$0.339 \pm 0.016$ GeV as well as a lattice determination $0.340 \pm 0.008$ GeV.  
Our value for the QCD string tension,  
$0.191\pm 0.009$ GeV$^2$ is also  in excellent agreement with the phenomenological value $\sigma \simeq 0.197$ GeV$^2$.
This connection between the fundamental hadronic scale underlying the physics of
quark confinement and the perturbative QCD scale controlling hard collisions can be carried out in any renormalization scheme.

We have also identified a scale $Q_0$  which defines 
the transition point between pQCD and nonperturbative QCD. Its value, $Q_0 \simeq 1$ GeV, is consistent with observations. 

\begin{acknowledgments} We thank Hans Guenter Dosch, Yang Ma, Xing-Gang Wu, and Xiaochao Zheng for valuable discussions. This material is based upon work supported by the U.S. Department of Energy, Office of Science, Office of Nuclear Physics under contract DE--AC05--06OR23177. This work is also supported by the Department of Energy  contract DE--AC02--76SF00515 (SLAC-PUB-16078). 
\end{acknowledgments}

\end{document}